# A Novel EM Gradiometric Surveying System for Geophysical Reconnaissance

Alexey V. Veryaskin[1,2], Francis A. Torres[2], Timo P. Vaalsta[1,2], Ju Li[2], David G. Blair[2]

Trinity Research Labs[1]
School of Physics[2], University of Western Australia, 35 Stirling Highway, Nedlands
Perth WA6009, Australia

## INTRODUCTION

In traditional audio-band electromagnetic (EM) deep-looking carrier-centred systems based on transmitter-target-receiver EM propagation chain, the in-phase component at the receiver station is normally orders of magnitude larger than the quadrature, which is 90 deg shifted with respect to the primary EM field (Lee, 2010). However, the latter component contains valuable information as it relates to the phase shift between the primary and secondary EM fields. As shown in (McNeill, 1980) the phase shift is directly proportional to the conductivity of the target in the so-called low induction limit and/or within the Extremely Low Frequency (ELF) band when the ability of a target to store EM energy can be neglected.

The appreciation of the information contained in the quadrature component of the secondary EM field resulted in the first frequency domain surveying systems built from the early 1950s to late 1970s. For example, the F-400 series quadrature airborne EM system is described in (Seiberl, 1974), which showed an equivalent phase noise of 0.5 ppt (part per thousand).

The first frequency domain systems, which were capable of measuring both the in-phase and quadrature components of the secondary field, were also being built. In order to cancel out the overwhelming disproportionality of the in-phase to quadrature EM components at the receiver station, some active compensation methods were used. For example, in the Barringer Research frequency domain system, the primary in-phase component was nulled out by a secondary transmitter, which was used to generate an EM field of opposite sign compared to the main transmitter.

In these early systems and in modern conductivity metres (or metal detectors), the quadrature demodulation principle is used (Horowitz and Hill, 1989). The same principle is used in lock-in amplifiers. In quadrature demodulators, the phase shift between the primary and secondary EM field is a result of calculating the ratio between the demodulated quadrature and the demodulated in-phase components of the secondary EM field, which are the measured quantities. The best equivalent phase noise performance that can be achieved in quadrature demodulation based devices is a few tens ppm (part per million), or about -100 dBc/√Hz, at the low-end of Very Low Frequency (VLF) audio band (see www.geonics.com).

In advanced phase sensitive interferometric systems, the inherent phase noise floor can be orders of magnitude lower compared to that of standard quadrature demodulation based systems (Ivanov *et al*, 1998; Rubiola, 2002). Below, a novel EM surveying system based on an interferometric principle and operating in the ELF audio-band is described. The ELF frequency range is most attractive in the case of surveying over highly conducting overburden. A gradiometric receiver configuration can also be used in this case.

## INTERFEROMETRIC PRINCIPLES APPLIED TO ELECTROMAGNETIC SURVEYING

There are a few different types of classic interferometric measurements. In Fig.1 below, the Mach-Zehnder type of RF & microwave interferometer is shown as an example (Ivanov *et al*, 1998). A single EM power carrier-centred source is used to pump in, through a power splitter, two arms of the interferometer. The measurement arm contains a Device Under Test (DUT) and the compensation arm contains precision balancing components, a phase shifter and an attenuator. These components are used for vector subtraction of the two EM signals propagating through each arm of the interferometer. A 4-port power combiner (CB) is used to redistribute the EM power in such a way that the output EM signal at the "dark port" (DP) of the interferometer is near zero. It is further amplified by a low noise amplifier (LNA) and mixed with a reference signal taken from the "bright port" (BP) of the power combiner. The reference signal is 90 degree phase shifted (by an additional phase shifter) compared to the matched signals in each arm of the interferometer. Any phase variations in the DUT will result in an imbalance of the matched arms of the interferometer and, therefore, in a voltage output directly proportional to the phase change in the DUT.

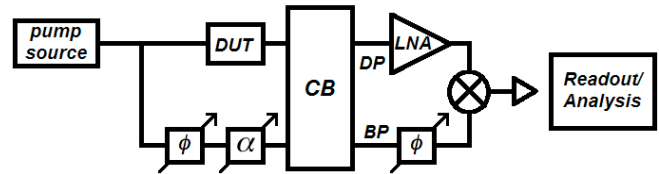

**Figure 1. The typical RF & microwave architecture of the Mach-Zehnder type of interferometer. A single frequency signal from a pump source is split in two signals that propagate through two arms of the interferometer. In the measurement arm (the upper arm) one passes through a Device Under Test (DUT). Another one, in the reference arm (the bottom arm) of the interferometer, passes through a precision phase shifter (f) and a precision attenuator (a). These components are used to balance the interferometer by making both the phase and the amplitude of the two signals equal with a high precision. As a result of the vector subtraction of the two signals at the power combiner station (CB) the EM power is near zero at the "dark port" (DP) of the interferometer. The carrier is heavily suppressed at this point so further amplification, by means of a low noise amplifier (LNA), is possible. The LNA output is then mixed with the reference signal coming through the "bright port" (BP) of the power combiner. If the reference signal is 90 degree phase shifted (by an additional phase shifter) before being mixed with the LNA output, any phase variation in the DUT will result in the mixer output signal directly proportional to the phase change in the DUT.**

In optical, microwave and RF interferometers it is desirable that both the measurement arm and the compensation arm of an interferometer have the same length. This is because the EM wavelength can be comparable with the size of the interferometer. In the audio band region the EM wavelength is many orders of magnitude larger than any practical transmitter-target-receiver EM propagation scale. This makes it possible to apply the Mach-Zehnder type of interferometric phase sensitive measurements to active source EM geophysical surveying (Veryaskin, 2010).

Below, in Fig.2, a block diagram of a practical realisation of an ELF EM interferometer (or phase bridge) adopted for EM surveying is presented. In this symmetric configuration both arms of the interferometer are measurement arms. The transmitter (Tx) plays the role of the EM power splitter and the target acts as a DUT as it introduces a phase shift between the primary (Bp) and the secondary (Bs) EM fields.

Each arm consists of a receiver (any low noise magnetic-field-to-voltage converter can serve to the purpose), a precision digital phase shifter, a hard limiter and a low pass filter. The low pass filter selects the first harmonic of the saturated signal at the output of the limiter. Its phase is the measure of the presence of the quadrature component of the secondary EM field at the corresponding receiver station. The limiter cancels amplitude variations of either signal in either arm of the interferometer. Another mechanism of seeing only phase variations between the primary and the secondary EM fields at the receiver stations is inherent to the mixer stage of the interferometer. It provides the phase-only sensitive output, which is used to maintain a closed loop operation that keeps the balance between the two arms of the interferometer unchanged. As depicted in Fig.2, a digital PI controller interface (DPI Ctr) is used to control a digital phase shifter, which compensates any phase imbalance between the two arms with a resolution of a few tens ppb (part per billion). The complimentary phase shifter is used for calibrating the interferometer output in the radian units.

The signals in either arm of the interferometer are combined at the combiner station and cancel each other if their phases and amplitudes are precisely matched. Typically one part in ten thousand of the phase balance is achieved, and one part in a hundred of the amplitude balance is enough to provide a reasonable "phase gain" in front of the traditional mixer stage.

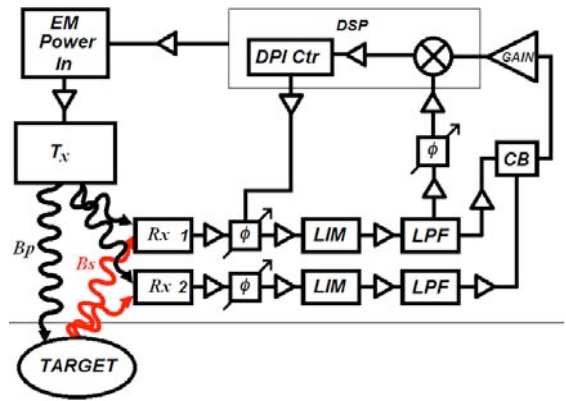

**Figure 2.** A symmetric realisation of the EM ELF interferometric system adopted for EM surveying. It is a close analogy to the Mach-Zehnder type of RF & microwave interferometer depicted in Fig. 1. An EM power source pumps energy into a transmitter (Tx), which acts as a power splitter. Direct EM coupling between the transmitter and the receivers (Rx1, Rx2) creates identical signals in either arm of the interferometer. Target changes the phase of either signal and is equivalent to a dual function DUT. Two hard limiters (LIM) in each arm of the interferometer are precisely matched and controlled by the same voltage reference. This removes the need of having a precision attenuator for the amplitude balance that is inherent to the classic (non symmetric) interferometer architecture shown in Fig.1. Low pass filters (LPF) select the first harmonic of the saturated signals, which is insensitive to any possible amplitude variations of the primary ($B_p$) and the secondary ($B_s$) EM fields at the receiver stations. The phase difference between the measurement arms of the interferometer at the combiner station (CB) is the measure of the presence of the quadrature component of the secondary EM field at the corresponding receiver station.

The interferometer inherent phase noise is inversely proportional to the EM power pumped into it (Ivanov *et al*, 1998). Therefore, it is desirable to build up the EM power at the receiver stations to a reasonably high level that provides the best noise performance. This situation is radically different from the traditional quadrature demodulation based systems, which can only deal with weak signals at the receiver stations. In fact, the direct transmitter-receiver coupling and the overwhelmingly large in-phase component of the secondary EM field (if any) both become positive factors in terms of providing the required EM power to either arm of the interferometer. As both arms are identical (the static dispersion imbalance error is typically one part in a few hundred), the frequency fluctuations in the carrier, which result in the direct frequency-to-phase-noise conversion, cancel each other at the output of the interferometer.

Another advantage of using the interferometric phase-sensitive detection in the ELF audio band is that the traditional mixer stage and the final signal processing stage (including the DPI controller interface) can both be shifted into a digital signal processor (DSP), which is a software based environment. This cannot be done in microwave and RF interferometers where the whole interferometer architecture is built of hardware.

If the receiver stations Rx1 and Rx2 are spatially separated and aligned along the same axis, the system becomes a phase gradiometer. However, a non-gradiometric configuration can easily be arranged if one of the receivers (say Rx1) is isolated from the secondary EM field. In this case the corresponding arm becomes the traditional reference arm of the interferometer.

## A NOVEL EM ELF GRADIOMETER

A fully deployable prototype of the EM ELF gradiometer has been built and tested in the field. The instrument's mobile transmitter-receiver section consists of two concentric transmitter coils (Tx1 and Tx2, see Fig. 3 below) mounted together on a solid non-conducting and non-magnetic concentric frame, with two pairs of induction coils (Rx1, Rx2, Rx3 and Rx4) separated spatially by a 0.5 m baseline. The coils are wound on high magnetic permeability cores with about 1 kHz operational corner frequency. They are firmly mounted inside the cylindrical space, occupied by the transmitter loops, in such a way that reduces to a minimum the effect of any possible mechanical motion of the receiver coils relative to the transmitter coils.

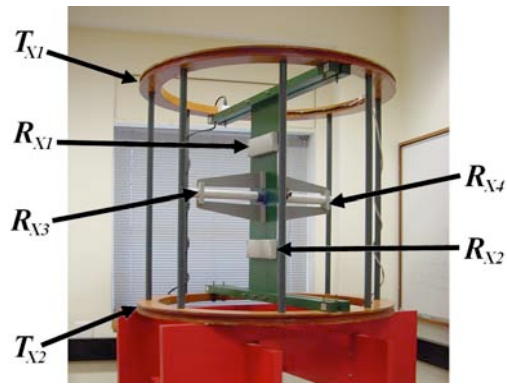

**Figure 3. The transmitter-receiver section of the EM ELF gradiometer. Tx1, Tx2 – transmitter loops wound on circular wooden frames with 1.4 metres in diameter, 100 turns each. The typical operation current is 10 A peak. Rx1, Rx2, Rx3, Rx4 – receiver coils.**

The system enables the detection of four AC gradients, namely Bxz, Byz, Bxx and Byy, dependent on the orientation of the transmitter-receiver section with respect to a coordinate reference frame. A classic (non symmetric) interferometer configuration was built and used in the prototype under test, similar to that of shown in Fig.1 (see www.trinitylabs.com.au).

The EM ELF gradiometer was tuned to operate at an ELF (~19 Hz) carrier frequency as it was intended for a field trial over an area with highly conducting overburden. During the field test, only one pair of the receiver coils (Rx1, Rx2), connected in a gradiometric configuration, was used. Their sensitivity axes were aligned along with the motion of a custom built trolley (with no metal parts), which was towed behind a 4WD vehicle loaded with the rest of the gradiometer hardware. The trolley-mounted transmitter-receiver section of the EM gradiometer is depicted in Fig.4.

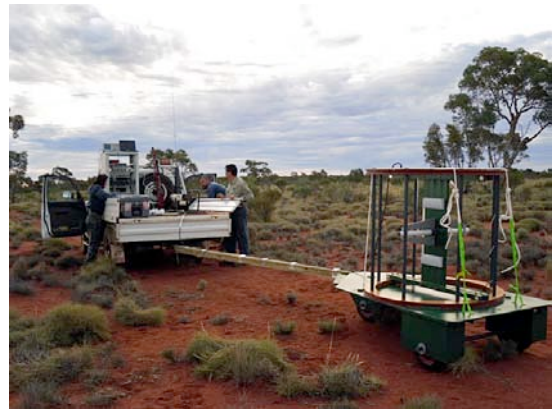

**Figure 4. The EM ELF gradiometer field trial with Regis Resources Pty Ltd commenced in July 2009.**

The monitoring system (CPU, oscilloscope, power amplifier, adjustable power supply, laptop) powered by a small generator was temporarily set up in the back of a Landcruiser trayback. This was connected to the transmitter-receiver unit, which was towed 5 metres behind the vehicle, mounted on the wooden trolley.

Initially the gradiometer survey was completed across several known EM conductors located at the field trial area. On most survey lines the data was collected using 2 methods:

1. readings were taken while the vehicle was run at a constant speed ~4 km/hour,

2. 10 second readings were collected at each station at 5m station spacings.

Both methods worked well if the surface was graded along the grid line. Where the grid line was rough the trolley-mounted transmitter-receiver section experienced vibration and the speed of safe travel was reduced. Five second reading times were tested, but were found to be too short, as the transmitter-receiver did not have enough time to stabilise. The condition of the track or grid line ultimately determined which method was most practical. Some runs were tested twice using the above methods to assess the repeatability of the data. Several known anomalies and a large EM conductor were easily identified.

The EM gradiometer was then trialled in an area with highly conductive overburden, including the Giles Nickel Prospect. The first line tested a known carbonaceous shale where an obvious target was detected. The system was then run across several lines at the Giles Nickel Prospect. A number of anomalies were discovered when the data were processed after completing the field trial. Previously, the area was surveyed with moving loop TEM (MLEM) using single turn 200m x 200m loops and base frequency 2.08 Hz. Analysis of the data showed a distinct lack of obvious deeper conductive features. An example of real time data recorded with the EM ELF gradiometer in an area with highly conducting overburden is shown in Fig.5.

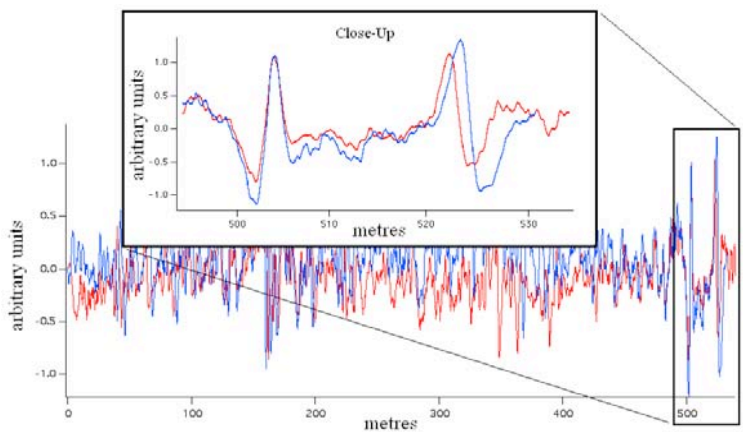

**Figure 5. Real time data recorded from the EM ELF gradiometer looping along the same line with a constant speed (~4 km/hour). Red and blue data sets show data from the same line but collected separately at different times. Strong correlation between the red and blue data sets indicates that real conductivity gradients were measured, not random noise.**

It is worth noting that signal drift was found to occur in the mornings, and was less severe in the afternoon. It was later discovered that the receivers were not completely covered and therefore exposed to sunlight in the morning. The transmitter-receiver unit would need to be completely protected from direct sunlight in order to avoid this problem.

Polarity of the output signal was found to be different on several lines and one was the reverse to another. Polarity across the Giles Nickel Prospect area remained constant indicating that the conductive overburden is consistent in this area.

## CONCLUSIONS

This paper describes a new ELF interferometric technique for the purpose of EM geophysical surveying. However, the full understanding of the potential of the novel EM ELF gradiometer is yet to be achieved and reported elsewhere. Especially, extensive modelling is required in order to convert raw data into fully informative commercial grade data sets. A newer version of the EM ELF gradiometer, enabling the more advanced symmetric interferometer architecture (see Fig.2), has been built and tested in the laboratory. The test results, a calibration strategy and the system performance will be published in due course.

The EM ELF gradiometer system can be deployed from a small helicopter or towed behind a 4WD vehicle of medium size in reasonably flat terrain. A towed version has been tested in regional outback Australia in relatively difficult terrain. The results of the field tests have been positive. The system was able to identify known EM conducting targets in the trial area and a number of anomalies were detected in the data collected in an area where geochemical signatures had indicated the likely potential of a massive nickel sulphide deposit.

## ACKNOWLEDGMENTS

Authors would like to thank Prof Eugeney Ivanov for his interest and helpful discussions in relation to advanced interferometric phase measurements. Authors would also like to thank Mr Howard Golden for his interest and overwhelming help in the process of developing the EM ELF technology and in writing this article. One of the authors (Alexey Veryaskin) would like to thank Mr Marc Kaye for his outstanding technical support in developing a commercial implementation of the EM ELF interferometric system. He also would like to thank Mr Jens Balkau, Ms Tara French and Mr Nathan Litchfield for an exciting field trial.